\documentclass{article}
\usepackage{graphicx}
\def\p{\partial}
\def\s{\sigma}

\def\g{\gamma}

\def\d{\delta}

\def\D{\Delta}
\def\ld{\lambda}
\def\Ld{\Lambda}
\def\L{\Lambda}
\def\ep{\epsilon}
\def\e{\eta}

\def\rh{\rho}

\def\b{\beta}

\def\a{\alpha}

\def\pdellx'{\frac{\partial}{\partial x'}}
\def\pdellw'{\frac{\partial}{\partial w'}}

\newcommand{\be}{\begin{equation}}
\newcommand{\ee}{\end{equation}}
\def\bed{\begin{displaymath}}
\def\eed{\end{displaymath}}
\def\bea{\begin{eqnarray}}
\def\eea{\end{eqncrray}}
\def\[{$$}
\def\]{$$}

\begin{document}
\title{A Unified Gravity-Electroweak Model \\ Based on
a Generalized Yang-Mills Framework}

\author{ Jong-Ping Hsu \\
Department of Physics,
University of Massachusetts Dartmouth \\
North Dartmouth, MA 02747-2300, USA}


\maketitle
{\small   Gravitational and electroweak interactions can be unified in analogy with the unification in the Weinberg-Salam theory.   The Yang-Mills framework is generalized to include space-time translational group $T(4)$, whose generators $T_{\mu}(=\p/\p x^{\mu})$ do not have constant matrix representations.  By gauging $T(4) \times SU(2) \times U(1)$ in flat space-time, we have a new tensor field $\phi_{\mu\nu}$ which universally couples to all particles and anti-particles with the same constant $g$, which has the dimension of length.  In this unified model, the $T(4)$ gauge symmetry dictates that all wave equations of fermions, massive bosons and the photon in flat space-time reduce to a Hamilton-Jacobi equation with the same `effective Riemann metric tensor' in the geometric-optics limit.    Consequently, the results are consistent with experiments.  We demonstrated that the $T(4)$ gravitational gauge field can be quantized in inertial frames. }
\bigskip

\bigskip

{\small PACS number: 04.20.Cv

\bigskip

The electromagnetic and weak interactions have been satisfactorily unified within the Yang-Mills framework with the internal groups $SU(2) \times U(1)$.  We demonstrate that the gravitational interaction of `Yang-Mills gravity,' which is based on space-time translational gauge symmetry,\cite{1}  can be similarly unified with the electroweak interactions of Weinberg-Salam theory.\cite{2}    For this unification involving gravity, we have to unify first external space-time gauge symmetry\cite{3} and internal gauge symmetry in a general framework, which may be called generalized Yang-Mills framework, for both inertial and non-inertial frames.\cite{4,5}  The translational symmetry in flat space-time (with
 vanishing Riemann-Christofell curvature) is particularly interesting because, through Noether's theorem, it is intimately related to the conservation of the energy-momentum tensor, which is related to the source of the gravitational field.   The generators of external four-dimensional space-time translational group $T(4)$ are denoted by $T_{\mu}$ in general, which satisfy $[T_{\mu}, T_{\nu}]=0$.  They have the representation $\p/\p x^{\mu} $ but do not have constant matrix representations, in sharp contrast to the generators of internal groups and the de Sitter group.\cite{6,7}   

Within such a unified gravity-electroweak (`gravielecweak')  model, we gauge $T(4) \times SU(2) \times U(1)$, which naturally allows us to introduce a new symmetric tensor gauge field $\phi_{\mu\nu}=\phi_{\nu\mu}$ through the replacement in the usual Yang-Mills framework,  
$$\p_{\mu} \to  \p_{\mu} + g\phi_{\mu}^{\nu}T_{\nu} -if{W_{\mu}^a}{ t^a} - if' U_{\mu}t_{o} \equiv d_{\mu} ,  \ \ \ \ \   c=\hbar=1, $$ 
where $t^a  \ ( a=1,2,3) $ are the generators of $SU(2)$ in general, and $t_o$ (weak hypercharge) is the generator of $U(1)$.\cite{7}   The operator $d_{\mu}$ may be called the unified gauge covariant derivative (UGCD).   There are three independent gauge coupling constants:  $f$ and $f'$ are dimensionless, and $g$ has the dimension of length, which is just right for gravity, because the $T(4)$ generator $T_{\mu}$ in general has the dimension of  (1/length)  in natural units.     A basic difference between gravity and electrodynamics is as follows:  The electron and the positron have the same attractive gravitational force characterized by one coupling constant $g$.  However, they have different electromagnetic forces (i.e., attractive and repulsive), which are characterized by two dimensionless coupling  constants $\pm {e}$.  This difference stems from the absence of an `$i$' in the $T(4)$ term involving $\phi_{\mu}^{\nu}T_{\nu}$ in the UGCD  $d_{\mu}$.  

In addition to the dynamical symmetry groups $SU(2)$ and $U(1)$, we postulate the local translational group $T(4)$ in flat space-time with the $T(4)$ gauge covariant derivative $  \p_{\mu} -i g\phi_{\mu}^{\nu}p_{\nu}=  \p_{\mu} + g\phi_{\mu}^{\nu}T_{\nu}$, where $T_{\nu}=\p_{\nu}=-ip_{\nu}$ are the $T(4)$ generators.   Thus, it is necessary and natural for a tensor field $\phi_{\mu}^{\nu}$ and the space-time translational generator  $T_{\mu}$ to appear together in the action through the unified gauge covariant derivative $d_{\mu}$.  In this sense, we have flat space-time translational origin of the gravitational field  $\phi_{\mu\nu}$ in the unified gravielecweak model, in contrast to other formulations of gravity. 

For simplicity, we assume the $T(4)$ tensor gauge field $\phi_{\mu\nu}$ to be symmetric for describing a spin-2 particle.  If one interprets the space-time translational group $T(4)$ to be basically the $R^4$ group, one will have a problem of consistence related to the symmetry property of gauge fields.  Suppose a gauge theory is formulated on the basis of the $R^4$ group, then one will have a set of 4 independent standard Abelian vector gauge fields rather than a tensor field $\phi_{\mu\nu}$.   The space-time translational group $T(4)$ differs from the group $R^4$ in their generators, associated gauge fields and gauge transformations.  For example, the $R^4$ gauge transformations look like 4 independent standard Abelian gauge transformations.  
However, the $T(4)$ gauge transformations in Yang-Mills gravity are not the usual gauge transformation which involve 4 independent  standard Abelian gauge transformations.   Rather, the T(4) gauge transformations are, for example, given by\cite{1}
$$
\d \phi_{\mu\nu}(x)= -\L^{\s}(x) \p_{\s} \phi_{\mu\nu}(x) - \phi_{\mu\s}(x)\p_{\nu} 
\L^{\s}(x) - \phi_{\s\nu}(x)\p_{\mu} \L^{\s}(x), 
$$
 where all indices are space-time indices.  The symmetry property $\phi_{\mu\nu} = \phi_{\nu\mu}$ is consistent with these $T(4)$ gauge transformations.  

In the formulation of Yang-Mills gravity,\cite{1} we did not start with 4 copies of four-vector gauge fields associated with the group $R^4$ and then combine them into one single tensor field $\phi_{\mu\nu}$.  Rather, the `tensor gauge field' $\phi_{\mu\nu}$ appears directly in the $T(4)$ gauge covariant derivative in equation (1), i.e., $  (\p_{\mu} + g \phi_{\mu\nu} T^{\nu} ) \psi  $, where $ T^{\nu}$ is $T(4)$ group generators $T^{\nu}=\p/\p x_{\nu}$ in Yang-Mills gravity.  Furthermore, if one uses the group $R^4$ instead of the group $T(4)$ in Yang-Mills gravity, one has dimensionless $R^4$ generators.  As a result, one will have a dimensionless coupling constant (for the interaction of $R^4$ gauge fields), which is wrong for the gravitational interaction.  Therefore, one cannot interpret the $T(4)$ group in Yang-Mills gravity to be basically the $R^4$ group.  (We shall briefly compare Yang-Mills gravity, teleparallel gravity and other gauge theories of gravity near the end of the paper.)

In this paper, we first demonstrate that the gauge covariant derivatives of Yang-Mills gravity and of Weinberg-Salam theory can be naturally combined to form the unified gauge covariant derivative (UGCD) $d_{\mu}$.  We derive the gauge curvatures associated with each of the gauge groups by  calculating the commutators of the UGCD $d_{\mu}$.  Next, we write down gauge invariant action involving quadratic gauge curvatures for both electroweak and gravitational interactions, as well as Higgs scalar and lepton fields.  We show that the wave equations of photon, leptons and scalar fields lead to the same effective Riemann metric tensor in the geometric-optics limit.  The results of the unified model are consistent with experiments and all  fields in  the model can be quantized.
  
Yang-Mills gravity and its $T(4)$ gauge invariance can be formulated in a general frame of reference (where inertial frames are the zero acceleration limit of non-inertial frames).\cite{1}  For simplicity, in the following discussions of gravielecweak model, its quantization and physical implications, we shall use inertial frames with $\e_{\mu\nu}=(+1,-1,-1,-1)$.  To see the gauge curvatures associated with each group in the UGCD $d_{\mu}$, let us calculate the commutators $[d_{\mu} ,d_{\nu}]$.   We obtain
\be
[d_{\mu} ,d_{\nu}]= C_{\mu\nu\a} T^{\a}  -ifW_{\mu\nu}^{a} t^a  - if' U_{\mu\nu} t_o,    \ \ \ \ \ \   [t^a, t^b]=i\ep^{abc}t^c,
\ee
$$d_{\mu}=J_{\mu}^{\nu}\p_{\nu}  -if{ W_{\mu}^a} t^a - if' U_{\mu}t_{o}, \ \ \ \ \ \ \ \    J_{\mu}^{\nu} = \d_{\mu}^{\nu} + g \phi_{\mu}^{\nu},$$ 
where $J_{\mu}^{\nu}\p_{\nu} \equiv \Delta_{\mu}$ may be called the $T(4)$ gauge covariant derivative.  
The $T(4)$ gauge curvature $C_{\mu\nu\a} $ in flat space-time  is found to be\cite{1}
\be
C_{\mu\nu\a}= J_{\mu\s}\p^{\s} J_{\nu\a}-J_{\nu\s} \p^{\s} J_{\mu\a}=\D_{\mu}J_{\nu\a}-\D_{\nu}J_{\mu\a}=-C_{\nu\mu\a},
 \ee 
which completely differs from the Riemann-Christoffel curvature in Einstein's gravity.  Fortunately, there exists a non-trivial gauge invariant action, which is  quadratic in new gauge curvatures $C_{\mu\nu\a}$  and is consistent with experiments.\cite{1}  (Cf. eqs. (4), (19) and (20) below.)
In the unified model, the new $SU(2)$ and $U(1)$ gauge curvatures in the presence of the gravitational gauge potential are, respectively, given by $W^{a}_{\mu\nu}$ and $U_{\mu\nu}$,
\be
W^{a}_{\mu\nu} =  J_{\mu}^{\s}\p_{\s}W^{a}_{\nu} - J_{\nu}^{\s}\p_{\s}W^{a}_{\mu}+ f \epsilon^{abc}W_{\mu}^{b}W_{\nu}^{c},  
\ee
$$
U_{\mu\nu}=  J_{\mu}^{\s}\p_{\s}U_{\nu} - J_{\nu}^{\s}\p_{\s}U_{\mu}=\D_{\mu}U_{\nu} - \D_{\nu}U_{\mu}.
$$
In the absence of gravity, i.e., $g=0$, the gauge curvatures in (3) reduce to those in the Weinberg-Salam theory.   We observe that the $T(4)$ gauge potential appears in $SU(2)$ and $U(1)$ gauge curvatures in (3), but the $SU(2)$ and $U(1)$ gauge potentials do not appear in the $T(4)$ gauge curvature in (2).  This is in harmony with the universal nature of the gravitational force, which acts on everything.  We also note that the gauge curvature of external space-time group and those of internal groups have different gauge transformation properties.\cite{1}  Nevertheless, the action of the gravielecweak model must be gauge invariant.

Based on gauge curvatures in (1)-(3), the new Lagrangian $L_{gew}$ of the unified gravielecweak model is assumed to take the form,
\be
L_{gew}=L_{\phi}+L_{WS},  
\ee
$$   
L_{\phi}= \frac{1}{4g^2}\left (C_{\mu\nu\a}C^{\mu\nu\a}- 2C_{\mu\a}^{ \ \ \  \a}C^{\mu\b}_{ \ \ \  \b} \right),$$
$$
L_{WS}= -\frac{1}{4}W^{a\mu\nu}W^{a}_{\mu\nu} -\frac{1}{4}U_{\mu\nu}U^{\mu\nu}  + \overline{L}i\g^{\mu}d_{\mu} L +\overline{R}i\g^{\mu}d_{\mu} R 
$$
$$
+(d_{\mu} \phi)^{\dagger} (d^{\mu} \phi) - V(\phi^{\dagger} \phi) -\frac{m}{\rho_o}(\overline{L}\phi R  +\overline{R}\phi^{\dagger} L),$$
 in the usual notations,\cite{7} i.e., R denotes the right-handed electron iso-singlet with only two independent spinor components, and L is the left-handed lepton iso-doublet, etc.  It is possible to include all the quarks and leptons in the three families\cite{7} without changing the UGCD in (1).  We observe that all terms in the original $SU(2) \times U(1)$ Lagrangian are affected by the presence of the gravitational field $\phi_{\mu\nu}$, except the last two terms of $L_{WS}$ in (4).  This property suggests that the spontaneous symmetry breaking and the masses of particles in the model are independent of the gravitational interaction. 
 
From the gravielecweak Lagrangian (4), we see that, similar to renormalizable Yang-Mills fields, the couplings of the gravitational  gauge field and electroweak gauge fields have no more than four-vertex in the Feynman diagrams.  The simplicity of $T(4)$ gauge symmetry for gravielecweak interactions paves the way for quantization of gravity\cite{8} and may offer a possible renormalizable quantum gravity, which deserves further investigation.
 
  To see the macroscopic experimental  implications of the gravielecweak model, it is more logical if one first derives the equation for the motion of a classical (or macroscopic) object from wave equations for quantum particles in the model, rather than making a separate assumption for the motion of classical objects.  Thus, it is interesting to explore the classical limit (i.e., the geometric-optics limit) of the wave equations for the photon, the electron, etc.  For bending of light by the sun, let us derive the eikonal equation for light ray from the new photon wave equation in the unified model.   As usual, we require that there is only one massless neutral gauge field in the unified model for the photon which is coupled to the electron with the charge $e(t_3 + t_o)$.\cite{7}  The Lagrangian involving the photon in the presence of gravity in this gravielecweak model is given by,
\be
L_{em} = -\frac{1}{4} F_{\mu\nu}F^{\mu\nu},
\ee
$$
F_{\mu\nu}= J_{\mu}^{\s}\p_{\s}A_{\nu} - J_{\nu}^{\s}\p_{\s}A_{\mu}, \ \ \ \ \ \
A^{\mu}= U^{\mu} cos \theta_{w}+W^{\mu}_3 sin \theta_{w},
$$
where the electromagnetic potential $A_{\mu}(x)$ is the usual combination of $U^{\mu}$ and $W^{\mu}_3$,\cite{7}  and $\theta_w$ is the Weinberg (or weak) angle.  This Lagrangian leads to a new photon wave equation in the presence of Yang-Mills gravity,
\be
\Delta_\mu F^{\mu\ld}  +
(\p_{\a}J^{\a}_{\mu}) F^{\mu\ld}=0, \ \ \ \  \ \ \ \  
\Delta_{\mu}=J_{\mu}^{\nu}\p_{\nu}.
\ee
Using the  gauge condition, $\p_{\mu}
A^{\mu}=0,$ and the limiting expression\cite{9} for the field
$A^{\ld}=a^{\ld}exp(i\Psi)$, we can derive the eikonal equation,
\be 
G^{\mu\nu}( \p_\mu \Psi) \p_\nu \Psi = 0, \ \ \ \ \ \ \  Ê
G^{\mu\nu} = \e_{\a\b} J^{\a\mu} J^{\b\nu}, 
\ee
in the geometric-optics limit (i.e., both the eikonal
$\Psi$ and the wave 4-vector $\p_{\mu}\Psi$ are very 
large).~\cite{9}  We stress that  only in 
 the geometric-optics limit (i.e., the classical limit) of the wave equation (6), an `effective Riemann metric tensor' 
 $G^{\mu\nu}$ emerges, as shown in (7). 

The Lagrangian for fermion (or electron) $\psi$ and Higgs field $\e$ in the presence of the gravitational gauge potential is given by
\be
L_{\psi}=
\frac{i}{2}[\overline{\psi} \g_\mu \Delta^\mu \psi - (\Delta^\mu
\overline{\psi}) \g_\mu  \psi] -
m\overline{\psi} \psi +\frac{1}{2}\D_{\mu} \e\D^{\mu}\e -\frac{1}{2}m_{H}^2 \e^2,
\ee
$$\g_{\mu}\g_{\nu}+\g_{\nu}\g_{\mu}=2\e_{\mu\nu},$$
where we have symmetrized the fermion Lagrangian.   The modified Dirac equation for the electron due to the presence of Yang-Mills gravity is
\be
i\g_\mu \Delta^\mu \psi - m \psi + \frac{i}{2} \g_{\mu}[\p 
_{\nu}(J^{\mu\nu} )] \psi = 0. 
\ee
Using the limiting expression for the fermion field
$\psi= \psi_{o}exp(iS)$, where $\psi_{o}$ is a constant spinor, and the properties that $g\phi_{\mu\nu}$ and $g\p_\a \phi_{\mu\nu}$ are extremely small for gravity, we can derive the fermion equation
\be
[\g_{\s} J^{\s\mu} \p_{\mu} S + m]\psi_o  = 0.     
\ee
In the classical limit, the momentum $\p_{\mu} S$ and mass $m$ are 
large quantities.
To eliminate the spin variables, as usual, we multiply a 
factor $(\g_{\ld} J^{\ld\nu} \p_{\nu} S - m)$ to  
 (10). 
With the help of the anti-commutation relation for $\g_{\s}$ in (8), equation (10) leads to the Hamilton-Jacobi 
equation,
\be
G^{\mu\nu}(\p_{\mu}S)(\p_{\nu}S) - m^{2} = 0,  \ \ \ \ \ \ \    G^{\mu\nu} = \e_{\a\b} J^{\a\mu} J^{\b\nu}, 
\ee
for the motion of a classical particle in the presence of the 
gravitational tensor field $\phi^{\mu\nu}$.  
 Similarly,  one can also verify that the equations for massive vector field $W_{\mu}^{a}$ in (4) and scalar field $\e(x)$ in (8) reduce to a Hamilton-Jacob equation of the form (11) in the geometric-optics limit.  

For experimental implications of the unified model based on (4), the Lagrangian $L_{WS}$ has been extensively discussed and experimentally confirmed.\cite{2,7}  So let us concentrate on the Lagrangian for $\phi_{\mu\nu}$ and a fermion field $\psi(x)$:
\be
L_{\phi\psi}= \frac{1}{2g^2}\left (\frac{1}{2}C_{\mu\nu\a}C^{\mu\nu\a}-
 C_{\mu\a}^{ \ \ \  \a}C^{\mu\b}_{ \ \ \  \b} \right)   + L_{gf} ,
\ee
$$+
\frac{i}{2}[\overline{\psi} \g_\mu \Delta^\mu \psi - (\Delta^\mu
\overline{\psi}) \g_\mu  \psi] -
m\overline{\psi} \psi,
$$
\be
L_{gf}=\frac{1}{2g^2}[\p_\mu J^{\mu\a} - \frac{1}{2}\p^\a J^\mu_\mu][\p^\nu 
J_{\nu\a} - \frac{1}{2}\p_\a J^\nu_\nu].
\ee
where $\Delta_{\mu} \equiv J_{\mu}^{\nu}\p_{\nu}$.  We have included a gauge-fixing term $L_{gf}$ to make the solution of gauge field equation well-defined.   The reasons are that field equations with gauge symmetry are not well defined in general and that it is a nuisance to find solutions of such field equations without using a gauge-fixing term or choosing a gauge condition.  

 The $T(4)$ gravitational field equation can be derived from (12),
\be
H^{\mu\nu} +  A^{\mu\nu}=  g^2 T^{\mu\nu},
\ee
$$
H^{\mu\nu} \equiv \p_\ld (J^{\ld}_\rho C^{\rho\mu\nu} - J^\ld_\a 
C^{\a\b}_{ \ \ \ \b}\e^{\mu\nu} + C^{\mu\b}_{ \ \ \ \b} J^{\nu\ld})  
$$
\be
- C^{\mu\a\b}\p^\nu J_{\a\b} + C^{\mu\b}_{ \ \ \ \b} \p^\nu J^\a_\a -
 C^{\ld \b}_{ \ \ \ \b}\p^\nu J^\mu _\ld,
\ee
\be
A^{\mu\nu} = \p^\mu \left(\p^\ld J_\ld{^\nu}  - \frac{1}{2} \p^\nu 
J^\ld_\ld \right)  - \frac{1}{2} \e^{\mu\nu} \left(\p^\a \p^\ld J_{\ld\a}  
 - \frac{1}{2} \p^\a \p_\a J^\ld_\ld \right), 
\ee
where $\mu$ and $\nu$ should be made symmetric in equations (14), (15) and (16).  When one uses arbitrary coordinates in flat space-time, the terms in $H^{\mu\nu}$ and $T^{\mu\nu}$ should be expressed in the covariant form, i.e., replacing ordinary partial derivatives by covariant partial derivatives.  However, the gauge-fixing term  $A^{\mu\nu}$ in (16) should stay the same (to break the translational gauge symmetry).   
  The source tensor $T^{\mu\nu}$ due to the fermion field is given by
\be
T^{\mu\nu} = \frac{1}{2}\left[ \overline{\psi} i\g^\mu \p^\nu \psi -
i(\p^\nu \overline{\psi}) \g^\mu \psi \right], 
\ee
Let us consider a static and spherically symmetric system, in
which  tensor gauge fields are produced by a spherical object at rest  with a mass $m$, i.e., $T_{00} = m\d^3({\bf r})$ and all other components of $T_{\mu\nu}$ vanish. 
Based on symmetry considerations, the 
non-vanishing components of exterior solutions $\phi^{\mu\nu}(r)$ are 
$\phi^{00}(r), \phi^{11}(r),  \phi^{22}(r)$ and 
$\phi^{33}(r)=\phi^{22}/ 
sin^2 \theta $, where $x^\mu =(w,r,\theta, \phi)$. To solve the static 
gauge potential field, we write
$
J^{00}= J^0_0  = S(r), \   -J^{11}= J^1_1= Q(r),$ and   $- r^{2}¥J^{22}= 
J^2_2= - r^2 \sin^2\theta J_{33}=J^3_3 = T(r).
$
The metric tensor is given by $ (1,   -1, -r^2, 
-r^2 \sin^2\theta )$.  

After a tedious but straightforward calculations,   we obtained the second-order approximation of the tensor field produced by the sun with mass $m$ in (14),\cite{1}
$$
g \phi^{00} = \frac{G m}{r} + \frac{G^2 m^2}{2r^2}, \ \ \ \ \
g \phi^{11} = \frac{G m}{r} + \frac{5 G^2 m^2}{2 r^2}, $$  
\be
g \phi^{22} = - \frac{1}{r^2} \left[-\frac{G m}{r} \right],  \ \ \ \ \  g \phi^{33} =\frac{ g \phi^{22}}{sin^2 \theta}, \ \ \ \ \ \   
\ee
where $G=g^2/(8 \pi)$ is the Newtonian constant.
  
We calculated the perihelion shift of the Mercury  to the second 
order with the help of $G^{\mu\nu}(r)$ in the Hamilton-Jacobi equation 
(11) in terms of the spherical coordinates.\cite{1,10}  
The advance of the perihelion for one revolution of the planet is found to be
\be
 \delta \phi \approx   \d \phi_{obs}\left(1 - 
\frac{3(E_o^2 - m_p^2)}{4 m_p^2} \right) ,  \ \ \ \ \ \   \d \phi_{obs}=\frac{6\pi m_{p}^2 G^2 m^2}{M^2}
\ee
where $m_{p}$, $E_o$ and $M$ are respectively the mass, the constant energy and the angular momentum of the planet.
We note that the second term ($\approx 10^{-7}$ for the Mercury) in the bracket of (19) shows a very small difference between  the result of gravielecweak model and the observation (or Einstein's gravity).~\cite{11,10}  
 
The bending of light by the sun can also be derived from the propagation of
a light ray in geometrical optics in an inertial frame.  Suppose the light
ray propagates in the
presence of the tensor gauge fields,  its path is determined by the
eikonal equation (7). 
Following the usual procedure,~\cite{1,10}  we find the following result 
for the deflection of a light ray,
\be
\Delta \phi \approx  \Delta \phi_{obs}\left(1 - 
\frac{18 G^{2}m^{2} 
\omega_o^{2}}{M^{2}} \right), \ \ \ \ \ \ \ \ \ \   \Delta \phi_{obs}=\frac{4Gm\omega_o}{M}
\ee
where $\omega_o$ denotes the frequency of the light.  The additional correction term ($\approx 10^{-8}$ for visible light) in the bracket differs from 
that of the observational result  $\Delta \phi_{obs}$  and is 
too small to be tested in the bending of light by the Sun.  
  
 With the help of geometrical optics,~\cite{1,12}   the red shift
can also be derived from  the  eikonal equation (7).
 The experiment of the time delay of radar echoes passing the sun 
can be explained  by the effective metric tensor $G^{11}(\rho)$  to the first order in $Gm/\rho$.~\cite{1,13}Ê
In gravielecweak model, the gravitational quadrupole radiations of 
binary pulsars can be 
calculated to the second-order in $g\phi^{\mu\nu}$ in inertial frames.  
The second-order result for the power emitted per solid angle in Yang-Mills 
gravity or gravielecweak model  can be shown to be consistent with that  obtained in Einstein's gravity.\cite{14,13}  

Although the gravielecweak model is based on gauge symmetry, however, in the geometric-optics limit, fermion and boson wave equations reduce to the same Hamilton-Jacobi equation (11), which involves `effective Riemann metric tensors' and has a strong analogy with the corresponding equation in Einstein's gravity.\cite{10}  
The limiting Hamilton-Jacobi equation (11)  can describe the motion of a free-fall classical object, the bending of light, and the perihelion shift of the Mercury in inertial frames.\cite{1}  Thus, in the geometric-optics limit, Yang-Mills gravity turns out to be compatible with the equivalence principle. 

For gauge transformations in the gravielecweak model, we assume that the tensor gauge field $\phi_{\mu}^{\nu}(x)$ is an electrically neutral field and an iso-scalar, so that the $T(4)$ gauge potential $\phi_{\mu}^{\nu}$  does not transform under  $SU(2)$ and $U(1)$ gauge transformations.  For other fields, $SU(2)$ and $U(1)$ gauge transformations  are the same as those in the Weinberg-Salam theory.\cite{7}   However, the $T(4)$ gauge transformations are more involved because the (infinitesimal) local and arbitrary space-time translations, 
\be
x'^{\mu} = x^{\mu}+\Ld^{\mu}(x),
\ee
are also  general coordinate transformations with arbitrary $\Ld^{\mu}(x)$.  Thus, the $T(4)$ gauge transformations for vector or tensor fields (in a general frame of reference with arbitrary coordinates within flat space-time) are formally the same as the Lie variation of tensors.\cite{1}  The action $S_{gew}$ with the Lagrangian (4) can be shown to be invariant under the $T(4)$ gauge transformation in general frames of reference, including inertial frames as limiting cases.
We may remark that space-time translational symmetry group $T(4)$ is the Abelian subgroup of the Poincar\'e group.\cite{15}   

The new viewpoint that the general coordinate transformations (21) can also be regarded as a local space-time translation provides an alternative approach to identify gravitational field as the $T(4)$ gauge field and to bring gravity into the generalized Yang-Mills framework in flat space-time.  Thus, Yang-Mills gravity can be quantized\cite{16} and, furthermore, unified with the electroweak theory. 

As usual, we assume that  the Faddeev-Popov method can be applied to quantize the gauge fields in the unified gravielecweak model.\cite{17,18}  Since the quantization of the electroweak sector is well-known in the Weinberg-Salam theory, let us concentrate on the quantization of the gravitational gauge field $\phi_{\mu\nu}$ and consider explicitly  the associated ghost fields (for gauge invariance and unitarity).\cite{17}   For this purpose, we choose the gauge-fixing terms $L_{gf}$ in equation (13) for $\phi_{\mu\nu}$.
Within an unimportant multiplicative factor, the vacuum-to-vacuum 
amplitude of quantum gravielecweak in inertial frames is
\be
W = \int d[Y^{\nu}] W(Y^{\ld})
exp\left[-i\int d^{4}x\left( \frac{1}{2g^{2}}Y^{\mu}Y_{\mu}\right)\right]
\ee
$$=\int d[\phi_{\a\b}, f_{WS}](det \ U) 
exp\left[i\int d^{4}x \{L_{\phi}+L_{WS} + L_{gf}\right],$$
$$
W(Y^{\ld})= \int d[\phi_{\a\b}, f_{WS}]
\left(exp\left[i\int d^{4}x(L_{\phi}+ L_{WS})\right] \right.$$
$$\left. \times (det \ U) \mbox{\large \boldmath $ \Pi$}_{x,\nu}\d(\p_{\mu}J^{\mu\nu}-\frac{1}{2}\p^{\nu}J-Y^{\nu})\right),
$$
where $Y_{\ld}= (1/2) (\d^{\mu}_{\ld}\p^{\nu} + \d^{\nu}_{\ld}\p^{\mu} - 
 \e^{\mu\nu}\p_{\ld})J_{\mu\nu},$ and $f_{WS}$ denotes all fields in the Weinberg-Salam theory.\cite{7}
As usual, the functional determinant $(det \ U)$ in (22) can be expressed in terms of an 
effective Lagrangian for ghost
vector-fermion fields $V^{\mu}(x)$ and $\overline{V}^{\mu}(x)$,
\be
det \ U=\int exp \left(i\int L_{eff} d^{4}x\right)
d[V(x)^{\ld},\overline{V}(x)^{\ld}],
\ee
$$
L_{eff}= \left\{ \frac{}{} \ \overline{V}^{\mu}(\e_{\mu\nu}\p_{\ld}\p^{\ld})V^{\nu}\right.
 -g (\p^{\ld} \overline{V}^{\mu}) 
(\p_{\nu}\phi_{\mu\ld}-\frac{1}{2}  
\eta_{\mu\ld}\p_{\nu}\phi)V^{\nu}
$$
$$\left.-g(\p_{\ld}\overline{V}^{\mu})[\phi_{\mu\nu}\p^{\ld}+ 
\phi_{\nu}^{\ld}\p_{\mu} -  
\d_{\mu}^{\ld} \phi_{\nu\s}\p^{\s}] V^{\nu}\frac{}{}  \right\}, $$
where   $\overline{V}^{\mu}$ 
is considered as an independent ghost field.
 Based on (22) and (23), the quantum gravielecweak is 
 given by the total Lagrangian $L_{tot}$,
\be
L_{tot}=L_{\phi}+L_{WS}+L_{gf}+ L_{eff},
\ee
where the Lagrangians $L_{\phi}$, etc.  are given by equations (4), (13) and (23).
    
Feynman rules for the gravielecweak model can be obtained from (24).  For example, the propagator of the ghost vector particle can be obtained as  
\be
G^{\mu\nu}=\frac{-i}{k^{2}} \e^{\mu\nu},
\ee
where the $i\ep$ prescription is understood.  The propagator of the graviton is 
\be
G_{\a\b,\rh\s}=\frac{-i}{k^{2}}(\e_{\a\b}\e_{\rh\s}-
\e_{\rh\a}\e_{\s\b}-\e_{\rh\b}\e_{\s\a})
\ee
$$
+\frac{3}{k^{4}}(k_{\s}k_{\b}\e_{\rh\a}
+k_{\a}k_{\s}\e_{\rh\b}+ k_{\rh}k_{\b}\e_{\s\a}+
k_{\rh}k_{\a}\e_{\s\b}).$$
The ghost-ghost-graviton vertex (denoted by 
$\overline{V}^{\mu}(p)V^{\nu}(q)\phi^{\a\b}(k)$) is found to be
\be
\frac{ig}{2}[p^{\a}k^{\mu}\eta^{\nu\b} + p^{\b}k^{\mu}\eta^{\nu\a} +
p\cdot q(\eta^{\mu\a}\eta^{\nu\b}+ 
\eta^{\mu\b}\eta^{\nu\a})+p^{\nu}q^{\b}\eta^{\mu\a}
\ee
$$ 
+ p^{\nu}q^{\a}\eta^{\mu\b}- p^{\nu}k^{\mu}\eta^{\a\b}
-  p^{\b}q^{\nu}\eta^{\mu\a} -  
p^{\a}q^{\nu}\eta^{\mu\b}].$$
We have used the convention that all momenta are incoming to the 
vertices.   The vector-fermion appears, 
by definition of the physical subspace, only in closed loops in the 
intermediate states of a physical process, and 
there is a factor of -1 for each vector-fermion loop in the unified model.  The Feynman rules for other propagators and vertices can also be obtained from (24).  For details of quantization and Feynman rules, we refer to reference 16.   

We observe that Yang-Mills gravity and quantum chromodynamics can also be unified in the generalized Yang-Mills framework.   Suppose one gauges the group $T(4) \times [SU(3)]_{color}$, one has quantum `chromogravity' with the gauge covariant derivative  
\be
Q_{\mu}=\p_{\mu} + g\phi_{\mu}^{\nu}T_{\nu} +ig_{s}{G_{\mu}^k}\frac{{ \ld^k}}{2},  
\ee 
where $\ld^k/2$ are the fundamental representation of the eight generators of color $SU(3)$.\cite{7}    Gauge curvatures and the action for such a model can be constructed from (28), which will be discussed in a separate paper. 

In the literature, there were many discussions about the possible connection between gravity and Poincar\'e group (and other groups).  Most of them followed Einstein's approach and based on curved space-time or using tetrads as dynamical variables.\cite{19}-[22]    Consequently, their Lagrangians differ from our $L_{\phi}$ in (4), and they have difficulties in the quantization of gravitational field.\cite{23}  One basic reason for the difference is as follows:  The `conventional gravity' is based on curved space-time, so that one has the Riemann-Christoffel curvature tensor to construct the invariant action.  On the contrary, Yang-Mills gravity is based on T(4) group in flat space-time, so that there is 
no Riemann-Christoffel curvature.   There is only T(4) gauge curvature $C^{\mu\nu\ld}$ to construct the gauge invariant action for gravity.  Moreover, the actions in the usual gauge theories and in Yang-Mills gravity involve quadratic gauge curvatures in flat space-time.   In contrast, the actions in the conventional formulations of gravity involve linear curvature of space-time.  Furthermore, we have the maximum 4-vertex for the self-coupling of gravitons in Yang-Mills gravity.  This  is much more simpler than that in Einstein's gravity, which involves N-vertex of graviton self-coupling, where N is an arbitrarily large number.\cite{16}  Thus, the present approach of Yang-Mills gravity is different from that in the conventional formulations of gravity.   

There is a formulation of gravity called teleparallel gravity (TG), which is based on the translational gauge symmetry on a flat space-time with a torsion tensor.\cite{24,25}   It is  closer to the Yang-Mills gravity in some aspects than the conventional formulations of gravity.   We note that in TG a flat connection with torsion makes translations local gauge symmetries and whose curvature tensor is the torsion field.  However, the concept of translational gauge symmetry is realized differently in Yang-Mills gravity and in TG.  Although there is a one-one correspondence between equations in Yang-Mills gravity and teleparallel gravity, they are really different.   To wit, the gauge covariant derivative in teleparallel gravity is defined by
$D_{\mu}\psi=\p_{\mu}\psi +B_{\mu}^{a}\p_{a} \psi = h^{a}_{\mu}\p_{a} \psi,$ where  $ h^{a}_{\mu} = \p_{\mu} x^a + B^a_{\mu}$.  Note that  $x^{\mu}$ is the coordinates of space-time with the metric tensor $g_{\mu\nu}(x)=\e_{ab}h^a_{\mu}(x) h^b_{\nu}(x)$, and  $x^{a}(x^{\mu})$ is the coordinates in a tangent Minkowski space-time with the metric tensor $\e_{ab}=(+,-,-,-).$  Thus, $h^{a}_{\mu}$ in TG is treated as a tetrad rather than a tensor.
The gauge curvature in TG is a torsion tensor $T^a_{\mu\nu}$:
$ [D_{\mu}, D_{\nu}]= T^a_{\mu\nu}\p_{a}, \ \ T^a_{\mu\nu}=\p_{\mu} h^a_{\nu}-  \p_{\nu} h^a_{\mu}$.  As a result, the the gravitational Lagrangian in TG\cite{24} differs from the Lagrangian $ L_{\phi}$ given by (4) in Yang-Mills gravity.  Thus, they are two different theories of gravity, even though they both involve translational gauge symmetry in flat space-time.
 
In conclusion, the gravitational interaction with
 space-time translational gauge symmetry is essential for its unification with the electroweak interaction.   It also assures the universal gravitational coupling to all particles in the unified gravielecweak model based on flat space-time.  Furthermore, the unification with translational gauge symmetry directly leads to an effective Riemann metric tensor $G^{\mu\nu} $ in the Hamilton-Jacobi equation  for the motion of  all particles in  the geometric-optics limit. 
Thus, it appears that a classical object moves in a `curved space-time' with the metric tensor $G^{\mu\nu}$.  However, the true underlying space-time of the physical world at the quantum level is flat, which allows for quantization of gravity by the Faddeev-Popov method.\cite{16,18} 
 The gravitational Lagrangian $L_{\phi}$ in (4) and the geometric-optics-limiting property of the model are the keys for  consistency with experiments (including the conservation of energy-momentum tensor).  A novel suggestion of the unified model is that the curvature of space-time revealed by classical experiments appears to be a manifestation of the flat space-time translational gauge symmetry in the classical limit.  

The work was supported in part by the Jing Shin Research Fund of 
UMassD Foundation.
The author would like to thank Leonardo Hsu for his collaboration  on the space-time symmetry in inertial and non-inertial frames for the generalized Yang-Mills framework.  He would also like to thank D. Fine for useful discussions.

\bibliographystyle{unsrt}

\end{document}